\documentclass[twocolumn,secnumarabic,amssymb, nobibnotes, aps, prl]{revtex4-2}
\usepackage{graphicx}
\usepackage{color}
\usepackage{amsmath}

\begin{document}
\title{Generation of high-fluence and high-intensity hard x-ray attosecond pulses  at  European XFEL}

\author{Ichiro Inoue$^{1,2,3}$} 
\email{ichiro.inoue@edu.k.u-tokyo.ac.jp}
\author{Ulrike Boesenberg$^{4}$}
\author{Rustam Rysov$^{4}$}
\author{Takahiro Sato$^{5}$}
\author{Ichika Harima$^{6,7}$}
\author{Chenzhi Xu$^4$}
\author{Jia Liu$^{4}$}
\author{Thomas M. Linker$^{5}$} 

\author{Zain Abhari$^{8}$}
\author{Andrei Benediktovitch$^{9}$}
\author{Uwe Bergmann$^{8}$}
\author{Ye Chen$^9$}
\author{Lu Cao$^9$}
\author{Winfried Decking$^9$}
\author{Gianluca Geloni$^4$}
\author{Marc Guetg$^9$}
\author{Trey Guest$^4$}
\author{Aliaksei Halavanau$^5$}
\author{J\"{o}rg Hallmann$^4$}
\author{Takashi Kimura$^{6,7}$}
\author{Naresh Kujala$^4$}
\author{Aliaksandr Leonau$^{4}$}
\author{Shan Liu$^9$}
\author{Tianyun Long$^9$}
\author{Johannes M\"{o}ller$^4$}
\author{Taito Osaka$^{2}$}
\author{Jan-Etienne Pudell$^{4}$}
\author{Weilun Qin$^9$}
\author{Nina Rohringer$^9$}
\author{Angel Rodriguez-Fernandez$^{4}$}
\author{Daniele Ronchetti$^{4}$}
\author{Yasuhisa Sano$^{2,10}$}
\author{Matthias Scholz$^9$}
\author{Roman Shayduk$^{4}$}
\author{Jergus Strucka$^4$}
\author{Yanwen Sun$^{5}$}
\author{James Wrigley$^4$}
\author{Kazuto Yamauchi$^{2,10}$}
\author{Diling Zhu$^{5}$}
\author{Alexey Zozulya$^{4}$}
\author{Makina Yabashi$^{2,11}$}
\author{Anders Madsen$^{4}$}
\author{Jumpei Yamada$^{2,10}$}
\email{yamada@prec.eng.osaka-u.ac.jp}
\author{Jiawei Yan$^{9}$}
\email{jiawei.yan@desy.de}

\affiliation
{
$^1$\mbox{Department of Advanced Materials Science, The University of Tokyo, Chiba 277-8561, Japan.}\\
$^2$\mbox{RIKEN SPring-8 Center, Hyogo 679-5148, Japan.}\\
$^3$\mbox{University of Hamburg, Institute for Experimental Physics/CFEL,  Hamburg 22761, Germany. }\\
$^4$\mbox{European X-Ray Free-Electron Laser Facility, Holzkoppel 4, 22869 Schenefeld, Germany.}\\
$^5$\mbox{SLAC National Accelerator Laboratory,  California 94025, USA.}\\
$^6$\mbox{Department of Applied Physics, The University of Tokyo, Tokyo 113-8656, Japan.}\\
$^7$\mbox{The Institute for Solid State Physics, The University of Tokyo, 5-1-5 Kashiwanoha, Kashiwa, Chiba 277-8581, Japan.}\\
$^8$\mbox{Department of Physics, University of Wisconsin-Madison, Madison, Wisconsin 53706, USA.}\\
$^9$\mbox{Deutsches Elektronen-Synchrotron DESY, Hamburg 22603, Germany.}\\
$^{10}$\mbox{Graduate School of Engineering, University of Osaka, Osaka 565-0871, Japan.}\\
$^{11}$\mbox{Japan Synchrotron Radiation Research Institute, Hyogo 679-5198, Japan.}\\
}

\begin{abstract}
By combining hard x-ray attosecond pulses from the European XFEL with total-reflection focusing x-ray optics, we generated nanofocused hard x-ray attosecond pulses with intensities and fluences comparable to the highest values attained in the hard x-ray regime.
A peak intensity on the order of 10$^{20}$ W/cm$^2$ is confirmed through the observation of saturation in amplified spontaneous emission from copper atoms.
These x-ray pulses enable new scientific opportunities, including the exploration of higher-order nonlinear light--matter interactions, damage-free structure determination, and coherent control of atoms and molecules.
\end{abstract}
\maketitle

Since the discovery of attosecond optical pulse generation via high-harmonic generation (HHG) \cite{McPherson1987, Ferray1988}, ultrafast measurements using attosecond optical pulses have revolutionized our understanding of electron dynamics on their natural timescales \cite{Krausz2009}. 
While HHG can generate attosecond pulses in the extreme-ultraviolet and soft x-ray regions, extending this approach to shorter wavelengths remains highly challenging because of the drastic reduction in conversion efficiency at higher photon energies \cite{Berrah:25}.

In the hard x-ray regime, x-ray free-electron lasers (XFELs) uniquely
combine atomic-scale wavelengths with high pulse energies and femtosecond
pulse durations \cite{McNeilNP2010}.
Recent advances in electron-beam manipulation have enabled FEL emission
from ultrashort slices of electron bunches, thereby shortening XFEL pulses
into the attosecond regime. Through several complementary approaches, attosecond FEL operation has been demonstrated in both the soft and hard x-ray regimes
\cite{HuangPRL2017, MarinelliAPL2017, Duris2020,
malyzhenkov2020single, prat2023coherent,
trebushinin2023, franz2024terawatt,
Yan2024, t8qq-h31p, funke2026, guo2026high}.
Hard x-ray attosecond pulses are expected to open new scientific opportunities. For example, their ultrashort duration allows x-ray measurements to be performed before substantial electronic excitation develops during pulse irradiation \cite{Inoue2021_atomic}. This capability may facilitate the exploration of higher-order nonlinear x-ray phenomena as well as the realization of damage-free x-ray diffraction and spectroscopy. Furthermore, multiple attosecond pulses generated using split-and-delay optics \cite{Yanwen2025} could enable coherent control of atoms and molecules through the interference of quantum pathways.

For many potential applications of  hard x-ray attosecond pulses, two photon parameters critically determine the measurement performance. The first is the fluence, defined as the pulse energy per unit area. High-fluence x-ray pulses are particularly advantageous for studies of microscopic samples
with dimensions smaller than the x-ray beam size.
The second is the peak intensity, defined as the fluence divided by the pulse duration. High-intensity x-ray pulses enable the induction of strong nonlinear effects.

Here, we achieve record-breaking fluence and peak intensity of  hard x-ray attosecond pulses by combining a high-power attosecond XFEL source with advanced focusing optics. The former is enabled by the self-chirping operation mode developed at the European XFEL facility \cite{Yan2024}.
In this scheme, an electron beam with a duration of a few femtoseconds and a high peak current ($\sim$10 kA) develops a strong energy chirp through longitudinal space-charge effects and coherent synchrotron radiation as it passes through the linear accelerator. 
The chirped beam is subsequently compressed in an arc section before being injected into the SASE-2 undulator, where it generates  hard x-ray attosecond pulses with pulse energies of a few hundred microjoules.
This  energy exceeds that of previously reported  hard x-ray attosecond pulses by more than an order of magnitude \cite{MarinelliAPL2017,HuangPRL2017,malyzhenkov2020single,t8qq-h31p}.
The focusing of x-ray pulses is achieved using a four-bounce total-reflection optical system, referred to as the advanced Kirkpatrick--Baez (AKB) mirror system. Owing to its achromatic response and high damage tolerance, the AKB system efficiently focuses the  hard x-ray pulses while preserving their temporal characteristics. 
Details of the optical design and performance are described elsewhere \cite{Inoue2025}.

\begin{figure}[t]
		\includegraphics[width=8.5cm]{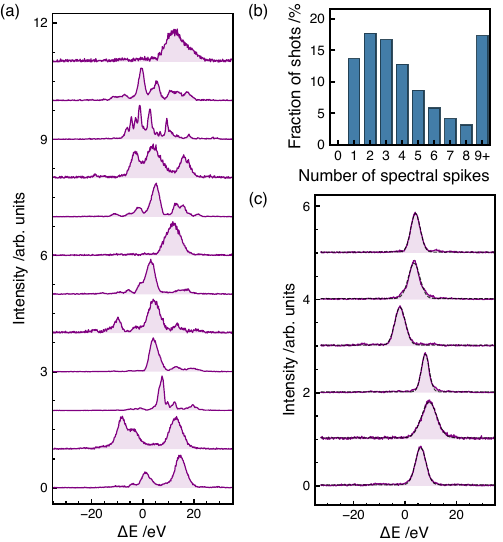}
		\caption{Spectral properties of attosecond x-ray pulses. (a) Single-shot spectra recorded over 12 consecutive pulses. (b) Histogram of the number of spectral spikes obtained from 74000 pulses. (c) Representative single-shot spectra of single-spike pulses. Dotted curves represent Gaussian fits to the spectra.}
\end{figure}

Experiments with focused attosecond hard x-ray pulses were carried out at the Materials Imaging and Dynamics (MID) instrument of the SASE-2 beamline at the European XFEL \cite{Madsen2021}.
The accelerator was operated in the self-chirping mode at an electron-beam energy of 14 GeV, generating attosecond x-ray pulses with a central photon energy of 10 keV at a repetition rate of 10 Hz. The pulse energy, measured by a calibrated x-ray gas monitor in the XTD1 tunnel \cite{Maltezopoulos2019} located 210 m downstream of the undulator exit, was 140 $\mu$J on average, with maximum pulse energy up to 380 $\mu$J.
Because of the long distance between the undulators and the experimental station (nearly 1 km), a beryllium compound refractive lens was inserted after the undulators to collimate the x-ray beam and match its size to the optical aperture downstream.
The beam was then transported by two flat mirrors coated with B$_4$C and operated at a grazing incidence angle of approximately 2.6 mrad  to suppress contamination from higher-order harmonics. The single-shot spectrum was measured non-invasively using the HIREX spectrometer, which combines a two-dimensional detector and a curved diamond (220) crystal and is located immediately downstream of the flat mirrors in the XTD6 tunnel \cite{Kujala2026}.
The AKB optical system consisted of two monolithic mirrors. Both mirrors were installed in the multipurpose chamber (MPC) of the MID instrument under vacuum conditions \cite{Madsen2021}. The two  mirrors were mounted on independent hexapods (Physik Instrumente).
A transmissive intensity monitor was positioned immediately upstream of the focusing optics. The monitor measured x-ray scattering from a 15-$\mu$m-thick polycrystalline diamond film (sp$^3$ Diamond Technologies) \cite{Tamasaku2016} using an ePix-100 detector \cite{Blaj2015}.
The  integrated scattering intensity for each pulse was correlated with the pulse energy at the focal point, which was measured using a laser power meter that had been calibrated beforehand against an x-ray calorimeter \cite{Kato2012}.

Figure 1a shows single-shot spectra recorded over 12 consecutive shots. 
The spectra typically consisted of only a few spikes, in marked contrast to conventional femtosecond XFEL pulses generated through self-amplified spontaneous emission \cite{Bonifacio1984, Huang2007}, whose spectra generally exhibit several tens of spikes \cite{Zhu2012, Inubushi2017, Kujala2026}. This spectral feature is consistent with lasing being confined to an ultrashort slice of the electron bunch. 
Figure 1b shows the distribution of the number of spectral spikes obtained from 74000 shots.
13.7\% of all pulses exhibited a single spectral spike. 
The spectrum of such single-spike pulses is nearly symmetric around the central photon energy and can be well approximated by Gaussian functions (Fig. 1c).

The pulse energy of these single-spike pulses at the focal point was 45 $\mu$J on average and up to 100 $\mu$J. 
Typical full-width at half-maximum (FWHM) bandwidth was 6 eV (Fig. 2(a)-(c)). 
Assuming a transform-limited Gaussian pulse, this bandwidth corresponds to a FWHM pulse duration of 300 as. The focal intensity distribution was characterized using the knife-edge scanning method with gold wires (Fig. 2d). In both horizontal and vertical directions, the beam profiles were described by  Gaussian functions, yielding FWHM focal spot sizes of 180 nm (horizontal) and 330 nm (vertical).

\begin{figure}[htb]
		\includegraphics[width=8.5cm]{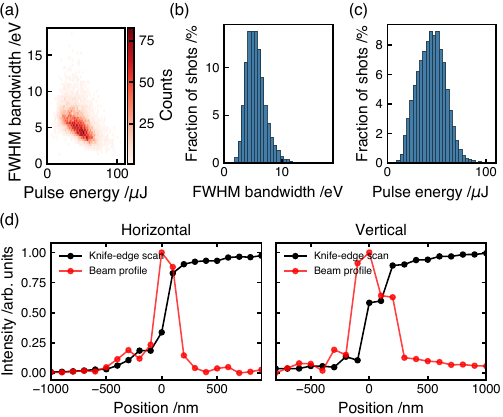}
		\caption{Beam properties of focused x-ray pulses with single spectral spikes. (a) Correlation between FWHM bandwidth and pulse energy for 10000 shots. (b) Histogram of the FWHM bandwidth. (c) Histogram of pulse energy. (d) Results of  knife-edge scan method (black) and evaluated intensity distribution of the focused beam (red).}
\end{figure}

Using the parameters given above, the fluence and peak intensity at the focal point were estimated to be 6.7 $\times$ 10$^{4}$ J/cm$^2$ and 2.1 $\times$ 10$^{20}$ W/cm$^2$, respectively. 
These values were calculated as $\frac{4 \ln 2}{\pi} \times$ 45 $\mu$J/(180 {nm}$\times$ 330 nm) for the fluence and  $(\frac{4 \ln 2}{\pi})^{3/2}\times$ 45 $\mu${J}/(180 nm $\times$ 330 nm $\times$ 300 as) for the peak intensity.
Here, the prefactors ($\frac{4 \ln 2}{\pi}$ and $(\frac{4 \ln 2}{\pi})^{3/2}$) accounts for the Gaussian spatial and temporal profiles defined by their FWHM beam sizes and pulse duration.
The achieved fluence and intensity are approximately one order of magnitude higher than those reported for attosecond XFEL pulses to date (2$\times$10$^3$ J/cm$^2$ and 3$\times$10$^{19}$ W/cm$^2$) \cite{Inoue2025}.

It is noteworthy that the achieved fluence and intensity are comparable to, or even exceed, the highest values reported so far for tightly focused femtosecond XFEL pulses \cite{Mimura2014, YumotoAS2020,Nagler2017,Yumoto2022}, with the exception of a very recent experiment that reached intensities on the order of 10$^{22}$ W/cm$^2$ \cite{Yamada2024}.
Such high-fluence and high-intensity femtosecond XFEL pulses have enabled a rich variety of applications,
including the exploration of nonlinear x-ray phenomena \cite{TamasakuNP2014,InouePRL2021,Inoue2023}, x-ray collective emission \cite{YonedaNature2015, Doyle2023, Linker2025, Kroll2026}, and high-resolution coherent imaging \cite{Yumoto2022,Yamaguchi2026}. 
The high-fluence and high-intensity hard x-ray attosecond pulses demonstrated in this study will extend these capabilities into the attosecond regime.

To verify the high peak intensity of the generated attosecond pulses, we investigated a nonlinear x-ray--matter interaction, namely amplified spontaneous emission (ASE) from copper atoms.
ASE is a collective radiative phenomenon in which spontaneously emitted photons are amplified through stimulated emission while propagating through a medium with population inversion. 
In the hard x-ray regime, ASE can be generated in 3$d$ transition-metal targets irradiated by high-intensity x-ray pulses with photon energies exceeding the $K$-shell binding energy \cite{YonedaNature2015,Linker2025}. 
The intensity threshold for the onset of ASE has been reported to be on the order of $10^{19}$ W/cm$^2$ \cite{YonedaNature2015}. 
Achieving saturation of ASE requires the creation of a sufficiently large population of core-hole states. 
One possible reason for ASE saturation is photoionization-induced depletion of the ground state on a timescale shorter than the fluorescence emission duration.
The fluorescence duration is limited by the core-hole lifetime, $\tau_c$,
while the photoionization rate can be approximated as $\sigma_{\mathrm{ph}}I/(\hbar\omega)$, 
where $\sigma_{\mathrm{ph}}$, $I$, and $\hbar\omega$ 
denote the inner-shell photoabsorption cross section per atom in the cold state, the x-ray intensity, and the photon energy, respectively. 
The condition for ground-state depletion during the fluorescence emission time then gives an estimate of the pump intensity required for ASE saturation as $\hbar\omega$ / $(\sigma_{\mathrm{ph}} \tau_c)$.
For copper irradiated with a 10-keV x-ray pulse, the intensity required for saturation of ASE is estimated to be $1.7\times10^{20}$ W/cm$^2$ using $\hbar\omega=10$ keV, $\sigma_{\mathrm{ph}}=2.3\times10^{4}$ barns/atom, and $\tau_c=0.4$ fs. 
Since this intensity is comparable to the estimated peak intensity of the present attosecond pulses, ASE provides an ideal benchmark for validating the generation of ultrahigh-intensity x-ray pulses.

ASE measurements were performed using a 20-$\mu$m-thick copper foil located at the focal position. The fluorescence signal emitted in the forward direction was measured with a Jungfrau 0.5M detector \cite{Mozzanica2018} positioned 600 mm downstream of the target. To prevent detector saturation and suppress the detection of transmitted and scattered 10-keV photons, an additional 120-$\mu$m-thick copper foil was placed 20 mm upstream of the detector.
Figure 3 shows the number of detected fluorescence photons as a function of the incident pulse energy at the target for single-spike pulses. The fluorescence yield exhibits an approximately exponential increase with pulse energy up to $\sim$20 $\mu$J, which is a characteristic signature of ASE just above the intensity threshold.
At higher pulse energies, the increase in fluorescence yield gradually slows and eventually saturates. The moving average of the detected fluorescence photon number approaches a constant value for pulse energies above 50 $\mu$J. 
This behavior is a clear signature for the saturation of ASE.
Assuming the pulse duration of 300 as, the peak intensity corresponding to the pulse energy of 50 $\mu$J is 
2.3 $\times 10^{20}$ W/cm$^2$. 
This value is in good agreement with the estimated intensity required to achieve ASE saturation (1.7 $\times 10^{20}$ W/cm$^2$), as discussed above. 
Therefore, this observation confirms generation of high-intensity pulses with intensity on the order of 10$^{20}$ W/cm$^2$.

\begin{figure}[!t]
		\includegraphics[width=7.5cm]{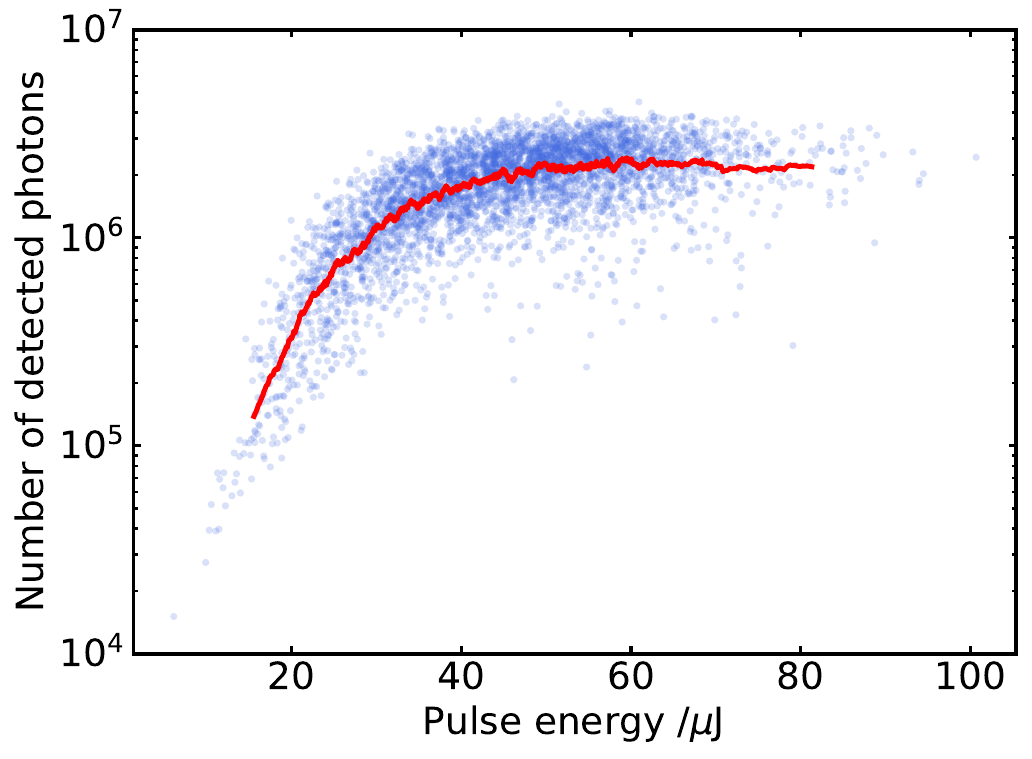}
		\caption{Number of detected fluorescence photons from 20-$\mu$m-thick copper foil irradiated with single-spike pulses. The red curve represents a moving average over 80 shots. }
\end{figure}

The hard x-ray attosecond pulses with high intensity and fluence demonstrated in this study extend the capabilities of x-ray free-electron lasers in many respects. 
For example, the demonstrated peak intensity highlights the potential of generated pulses as a platform for exploring higher-order nonlinear x-ray processes.
Beyond nonlinear x-ray science, these pulses also provide new opportunities for electronically damage-free structure determination by probing matter before substantial electronic relaxation and damage occur. 
Furthermore, the combination of ultrashort duration, high photon flux, and strong electric fields opens prospects for manipulating electronic wave packets and coherently controlling atomic and molecular dynamics with x-rays.
The demonstrated hard x-ray attosecond pulses thus mark a significant step toward observing, driving, and controlling electronic processes in matter on their intrinsic timescales with atomic-scale spatial resolution.

\section{Acknowledgements}
We acknowledge European XFEL in Schenefeld, Germany, for provision of x-ray free-electron laser beamtime at MID (proposal numbers of 7859 and 10568) and would like to thank the staff for their assistance, in particular Gabriele Ansaldi, Andreas Schmidt and Lennart Oppelt. We thank Kenji Tamasaku, Stephan Kuschel, Robert Radloff, Rasmus Bunchen, Jan Niklas Leutloff, Taisia Gorkhover, Lise Joost St{\o}ckler, Bo Iversen, Tomoki Fujita, Kiyofumi Takaba, Kensuke Tono, Yuichi Inubushi, Gota Yamaguchi, Jan Gr\"unert, Haixiao Deng, and Thomas Feurer for the fruitful discussion.

\section*{Data availability}
All data in this study are available from the corresponding author on reasonable request. Data recorded for the experiment at the European XFEL are available at doi.org/10.22003/XFEL.EU-DATA-010568-00.

\section*{Funding}
This research is supported by JSPS KAKENHI (Grant Numbers: 22KK0233, 23K17149, 23K25131, 24K21199) and JST Precursory Research for Embryonic Science and Technology (JPMJPR24J1). 

\section*{Competing interests}
The authors declare no competing interests.

\bibstyle{natbib}
\bibliography{Ref}

\end{document}